\documentclass[twocolumn]{aastex62}

\newcommand{\hi} {{\rm H\,{\footnotesize\rm I}}}
\usepackage{hyperref}
\usepackage{xcolor}

\shorttitle{The halo mass function of late-type galaxies}
\shortauthors{Li et al.}

\begin{document}

\title{The halo mass function of late-type galaxies from \hi\ kinematics}

\correspondingauthor{Pengfei Li}
\email{PengfeiLi0606@gmail.com, pxl283@case.edu}
\author[0000-0002-6707-2581]{Pengfei Li}
\affil{Department of Astronomy, Case Western Reserve University, Cleveland, OH 44106, USA}
\author{Federico Lelli}
\altaffiliation{ESO fellow}
\affiliation{European Southern Observatory, Karl-Schwarschild-Strasse 2, Garching bei M\"{u}nchen, Germany}
\author{Stacy McGaugh}
\affiliation{Department of Astronomy, Case Western Reserve University, Cleveland, OH 44106, USA}

\author{Marcel S. Pawlowski}
\affiliation{Leibniz-Institut f\"ur Astrophysik Potsdam (AIP), An der Sternwarte 16, D-14482 Potsdam, Germany}

\author{Martin A. Zwaan}
\affiliation{European Southern Observatory, Karl-Schwarschild-Strasse 2, Garching bei M\"{u}nchen, Germany}
\author{James Schombert}
\affiliation{Department of Physics, University of Oregon, Eugene, OR 97403, USA}

\begin{abstract}

We present an empirical method to measure the halo mass function (HMF) of galaxies. We determine the relation between the \hi\ line-width from single-dish observations and the dark matter halo mass ($M_{200}$) inferred from rotation curve fits in the SPARC database, then we apply this relation to galaxies from the \hi\ Parkes All Sky Survey (HIPASS) to derive the HMF. This empirical HMF is well fit by a Schecther function, and matches that expected in $\Lambda$CDM over the range $10^{10.5} < M_{200} < 10^{12}\;\mathrm{M}_{\odot}$. More massive halos must be poor in neutral gas to maintain consistency with the power law predicted by $\Lambda$CDM. We detect no discrepancy at low masses. The lowest halo mass probed by HIPASS, however, is just greater than the mass scale where the Local Group missing satellite problem sets in. The integrated mass density associated with the dark matter halos of \hi-detected galaxies sums to $\Omega_{\rm m,gal} \approx 0.03$ over the probed mass range.

\end{abstract}


\section{Introduction} \label{sec:intro}

The standard $\Lambda$ Cold Dark Matter ($\Lambda$CDM) model predicts the abundance of dark matter (DM) halos, which is quantified by the halo mass function (HMF) $\psi(M_{\rm halo})$, i.e., the number density of halos at a given halo mass. The analytic prediction \citep{PressSchechter1974} for $\psi(M_{\rm halo})$ is reproduced by N-body simulations of structure formation \citep{Warren2006, Boylan-Kolchin2009}. However, it is a challenge to compare the predicted HMF to observations since halo masses are hard to measure for individual galaxies, much less for a large sample.

Quantities accessible to observation include the luminosity and velocity functions of galaxies. These quantify the number density of galaxies as a function of luminosity and rotation speed, respectively. By adopting some prescription to estimate the mass-to-light ratios of stellar populations, the luminosity function can be transformed into the Stellar Mass Function (SMF). A simple comparison between the observed SMF and the $\Lambda$CDM prediction can be made by scaling the HMF by the cosmic baryonic fraction $f_b \approx 0.15$. This reveals a discrepancy at both high and low masses: the predicted HMF is a power law (since $\Lambda$CDM is scale-free), while the observed SMF is a Schecter function with a characteristic scale at $M_\star\simeq10^{10.5}$ M$_\odot$. This implies a non-linear variation of the stellar mass with halo mass that is attributed to feedback processes \citep{Bullock2017}. Abundance matching \citep[e.g.,][]{Behroozi2010, Moster2013} quantifies this variation by requiring a correspondence between the observed number density of galaxies and the expected number density of dark matter halos as a function of mass.

An independent approach is to consider the velocity function (VF) of galaxies, which probes more directly the galaxy potential well. Theoretically, the VF of galaxies can be constructed considering the maximum rotation velocity of DM halos ($V^{\rm DM}_{\rm max}$). Observationally, blind \hi\ surveys with single-dish radio telescopes provide the spatially integrated \hi\ line-width ($W_\hi$), which is a proxy for twice the rotation velocity of galaxies. The VF from \hi\ surveys is well-described by a modified Schechter function and differs from the one predicted in $\Lambda$CDM via $V^{\rm DM}_{\rm max}$ \citep[e.g.,][]{Zwaan2010, Papastergis2011} with possible implications for cosmology and the nature of DM \citep{Zavala2009, Klypin2015, Schneider2017, Schneider2018}. The comparison between theory and observations, however, is complex because the relation between $V^{\rm DM}_{\rm max}$ and $W_\hi$ may be strongly non-linear \citep[e.g.,][]{Brook2016, Maccio2016, Brooks2017, Chauhan2019, Dutton2019}.

In this letter, we present a new empirical method to directly measure the HMF of galaxies. We use 168 late-type galaxies from the Spitzer Photometry \& Accurate Rotation Curves (SPARC) database \citep{SPARC} to determine the relation between the \hi\ line width from single-dish observations and the halo mass from rotation-curve fits. This provides a tool to estimate halo masses from \hi\ line widths, and thereby translate the VF into the HMF. We apply this method to galaxies from the \hi\ Parkes All Sky Survey (HIPASS) catalogue \citep{Meyer2004} and provide the first direct comparison between the predicted and measured HMFs.

\section{Data}
\subsection{The HIPASS Galaxy Sample}

We use the sample of 1388 late-type galaxies with optical IDs and inclination larger than 45$^{\circ}$ \citep{Zwaan2010} selected from the \hi\ Parkes All Sky Survey (HIPASS) galaxy catalogue \citep{Meyer2004}. \citet{Zwaan2004} show that the completeness of this sample is 99\% at a peak flux of 84 mJy and at an integrated flux of 9.4 Jy km s$^{-1}$. This enables the measurement of galaxy abundance once the volume correction is appropriately taken into account. \cite{Zwaan2010} use these data to measure the VF. We utilize these same data to measure the HMF, using an effective conversion between \hi\ line width and DM halo mass.

\begin{figure*}
\epsscale{1.17}
\includegraphics[scale=0.31]{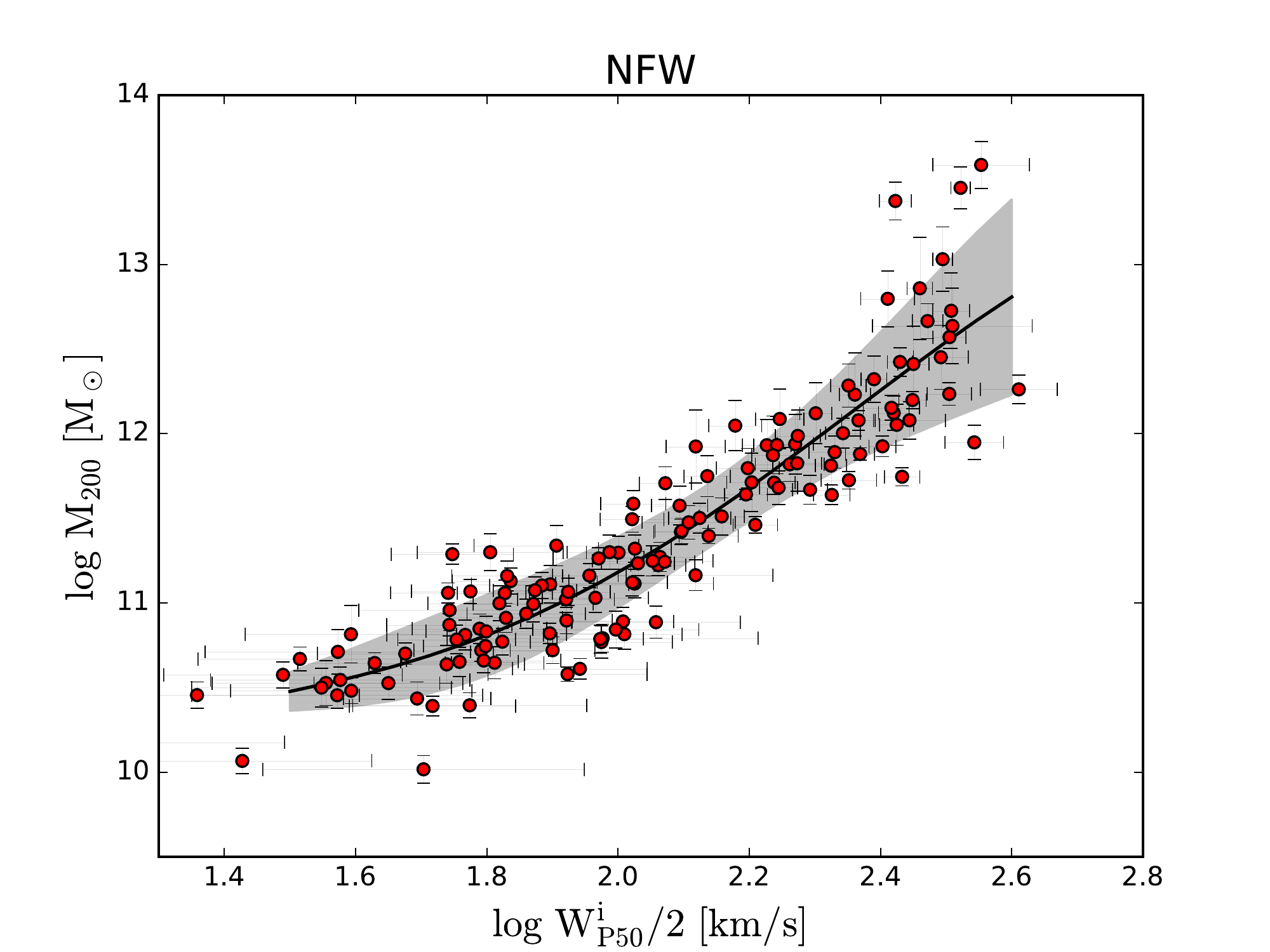}\includegraphics[scale=0.31]{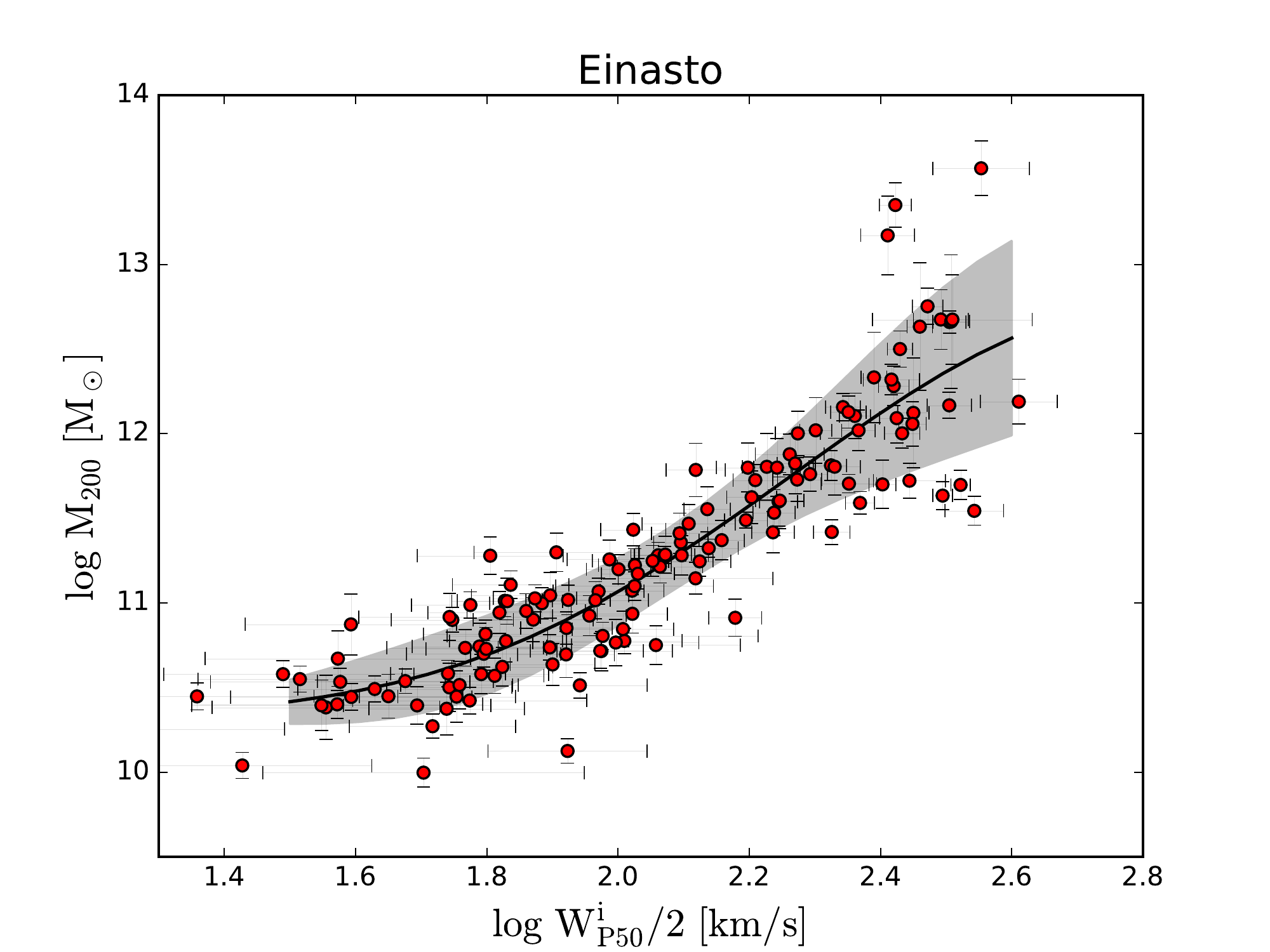}\includegraphics[scale=0.31]{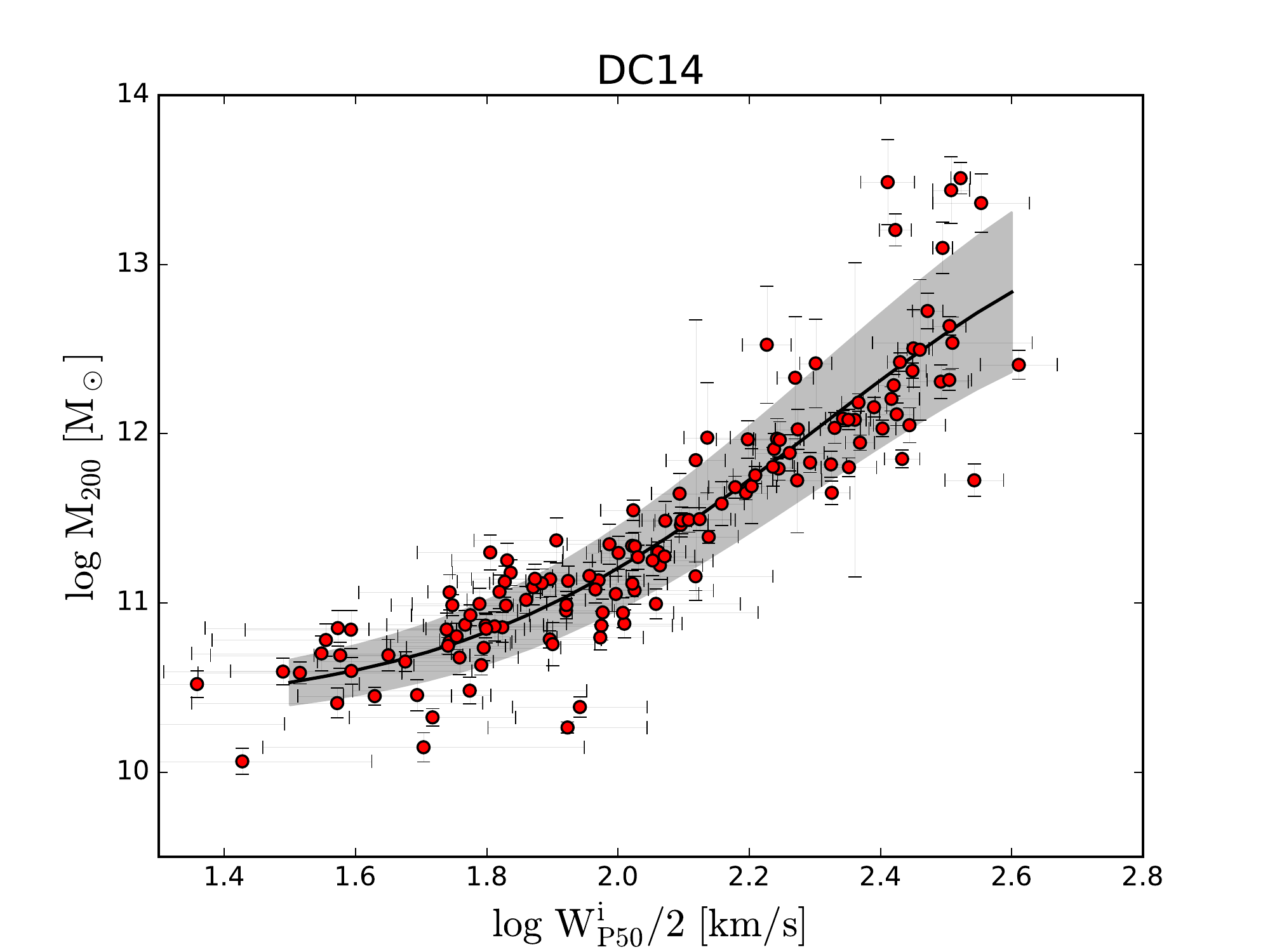}
\caption{Correlations between halo mass, $M_{200}$, and inclination corrected \hi-line widths, $W^i_{\rm P50}/2$, for SPARC galaxies. Halo masses are calculated from rotation-curve fits using the NFW (left), Einasto (middle), and DC14 (right) profiles. Solid lines are the best fits using the Gaussian Process Regression (GPR) algorithm and the shaded regions represent the GPR smoothed standard deviations.
}
\label{WP50M200}
\end{figure*}
\subsection{SPARC Rotation Curve Fits}

The SPARC sample \citep{SPARC} has measurements of rotation curves from spatially resolved interferometric data as well as \hi\ line widths spatially unresolved single-dish observations \citep{Lelli2019}. It includes 175 late-type galaxies with \hi/H$\alpha$ rotation curves traced to large radii, which constrain galaxy dynamical masses. 
This provides a way to explore the correlation between \hi\ line width and DM halo mass. 

\citet{Li2019} fit SPARC rotation curves using two simulation-motivated halo profiles, the Einasto \citep{Einasto1965, Navarro2004} and DC14 \citep{DC2014} profiles. 
These fits provide an estimate of the halo mass $M_{200}$ defined at the mass enclosed within an overdensity 200 times the critical density of the Universe. 
The fits were made imposing as priors the $\Lambda$CDM halo mass--concentration relation \citep{DuttonMaccio2014} and the stellar mass--halo mass relation \citep{Moster2013}. We discuss the role of the latter in section \ref{sec:disc}.

For reference, we also fit the commonly used NFW profile \citep{NFW1996} and derive halo masses following the same procedure, although it is well known that the NFW profile does not provide satisfactory fits to the rotation curves \citep{Katz2017}. The halo masses for the NFW profile thereby are less reliable than for the other profiles.

\subsection{The Single-Dish \hi\ Line Widths}

The \hi\ line widths for the SPARC galaxies are collected by \citet{Lelli2019}, mainly from the Extragalactic Distance Database \citep{Tully2009} but also from other references \citep[e.g.,][]{Springob2005, Huchtmeier1989}. In total, 168 out of 175 galaxies have the line-width measurements at 20\% of the peak flux density, i.e., $W_{\rm P20}$. To translate $W_{\rm P20}$ to the $W_{\rm P50}$ used by the HIPASS team \citep{Zwaan2010}, we adopt the conversion established by \cite{Courtois2009},
\begin{equation}
W_{\rm P50} = W_{\rm P20} - 26\ {\rm km/s}.
\end{equation}
This relation has an rms scatter of 21 km/s, which we propagate into the uncertainty in $W_{\rm P50}$. Although $W_{\rm P20}$ is also available in the HIPASS survey, \citet{Zwaan2010} use $W_{\rm P50}$ because it is less sensitive to noise in the \hi\ spectra. Thus, we adopt the same approach of \citet{Zwaan2010} for the HIPASS galaxies and simply convert $W_{\rm P20}$ into $W_{\rm P50}$ for the SPARC galaxies.

The measured line widths are projected along the line of sight. To recover the intrinsic widths, one has to correct the measurements for inclinations via $W^i_{\rm P50} = {W_{\rm P50}}/{\sin i}$. Optically defined inclinations have been extensively used for this purpose, since single-dish surveys cannot resolve the \hi\ distribution. Following the standard procedure \citep{Zwaan2010}, we calculate optical inclinations for the SPARC galaxies according to
\begin{equation}
\cos^2 i = \frac{q^2-q_0^2}{1-q_0},
\end{equation}
where $q$ is the axial ratio and $q_0=0.2$ accounts for the thickness of stellar disks. We measure the axial ratio from the outer isophotes of the [3.6] images based on those ellipses whose values differ from their mean by less than 20\%. SPARC galaxies have well measured kinematic inclinations, but we use the optical inclinations for internal consistency with HIPASS. The results are insensitive to the choice of which inclination we use.

\section{Results}
\subsection{The Halo Mass--Line Width Correlation}

In Figure \ref{WP50M200}, we plot halo mass, $M_{200}$, against line width, $W^i_{\rm P50}$/2. A strong correlation between $M_{200}$ and $W_{\rm P50}^i/2$ is apparent for each halo model. We use the Gaussian Process Regression (GPR) from the open python package $scikit-learn$ \citep{scikit-learn} to capture the mean relation (solid lines in Figure \ref{WP50M200}). The shaded areas show the estimated standard deviations smoothed by the GPR algorithm. 

This correlation has a well understood physical background. Roughly speaking, the inclination-corrected \hi\ line widths correspond to twice the rotation velocities since the SPARC galaxies are rotationally supported. The rotation velocity in the outer galaxy regions is mostly driven by the DM halo, thus one expects a correlation between $W_{\rm 50}$ and $M_{200}$. We can thus assign a halo mass to galaxies based on their much more readily measured line width. This enables us to map the HIPASS VF into any variable that correlates with line width.

\begin{figure*}[t]
\epsscale{1.17}
\plottwo{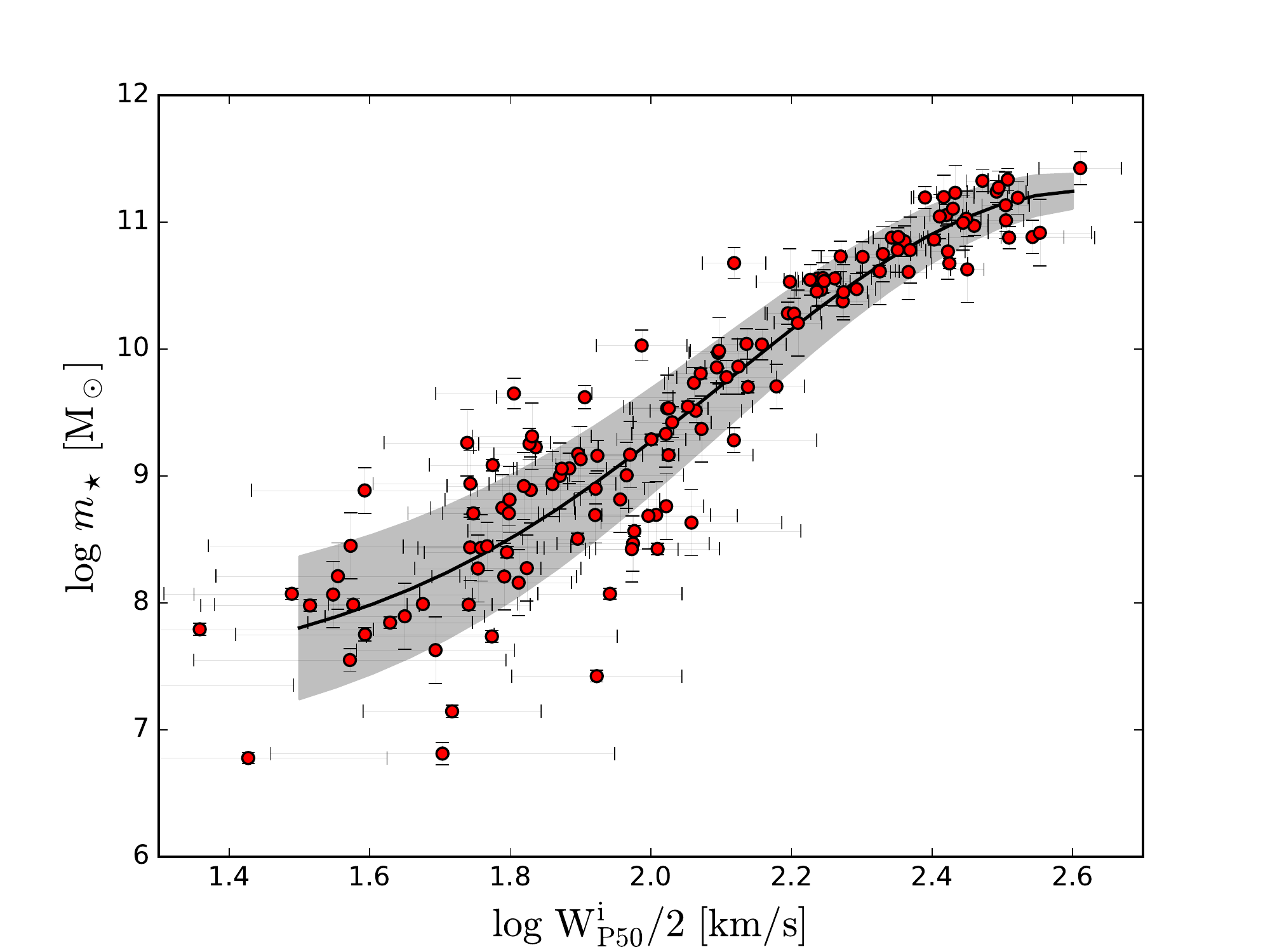}{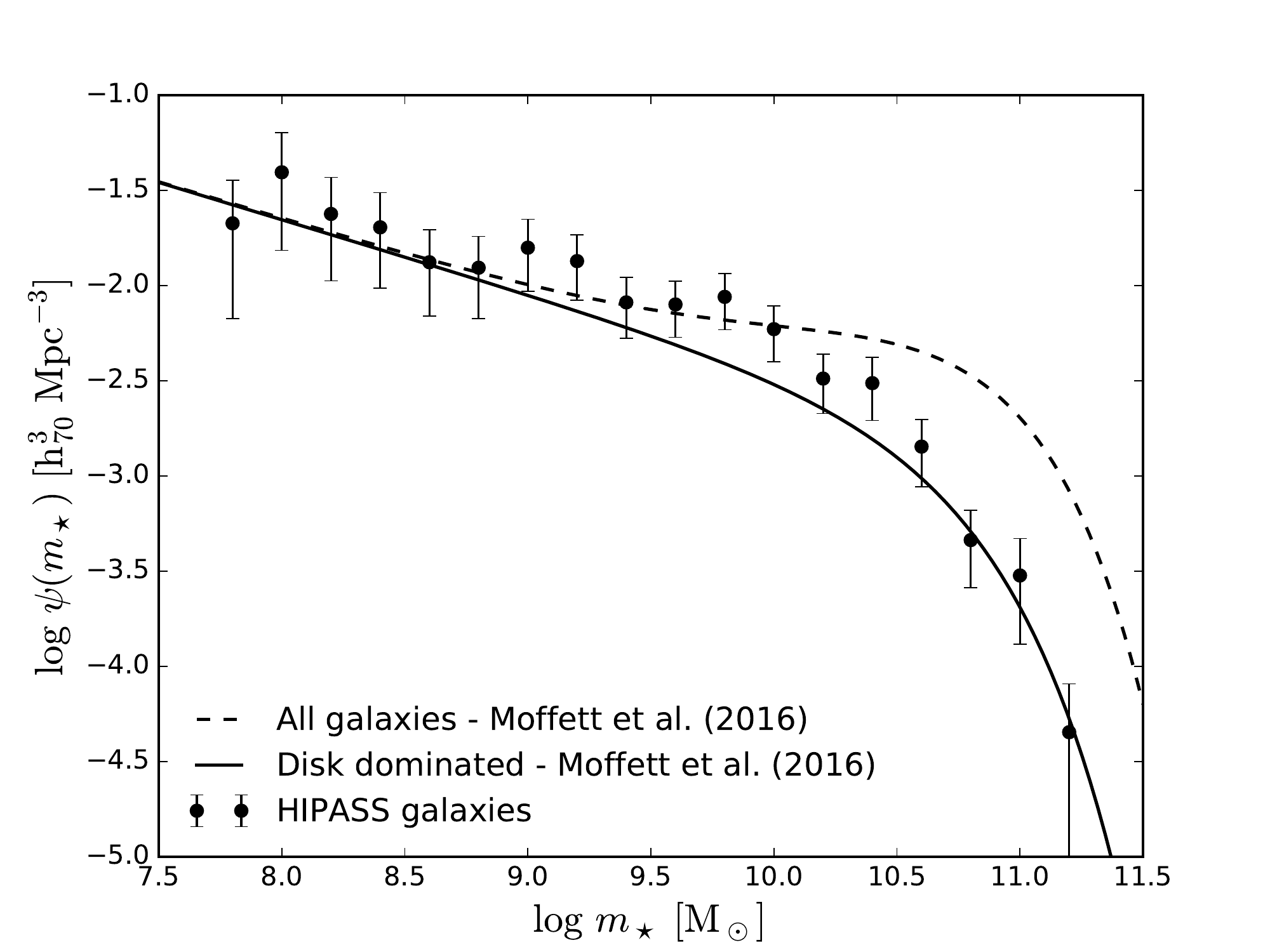}
\caption{Left: same as Figure \ref{WP50M200}, but for the stellar masses of SPARC galaxies assuming $\Upsilon_{\rm disk}=0.5$ and $\Upsilon_{\rm bul}=0.7$. Right: the stellar mass function (points) measured by applying our method to HIPASS galaxies \citep{Meyer2004, Zwaan2010}. The stellar mass function for galaxies in the GAMA survey \citep[][]{Moffett2016} is shown as the dashed line. This includes gas poor early type galaxies; the SMF of disk type galaxies (solid line) is a better match to the SMF we derive from HIPASS, as expected.}
\label{SMF}
\end{figure*}

\subsection{Stellar Mass Function}

To validate our method, we first derive the stellar mass function, which can be directly checked using the extensive measurements made with optical surveys \citep[e.g.,][]{Moffett2016,Wright2017,Jones2018}. To calculate the stellar masses of the SPARC galaxies, we adopt as fiducial values the [3.6] stellar mass-to-light ratios $\Upsilon_{\rm disk}=0.5$ and $\Upsilon_{\rm bul} = 0.7$ \citep{McGaugh2016PRL}. The SPARC galaxies show a strong correlation between $\log m_\star$ and $\log W^i_{\rm P50}/2$ as expected from the \citet{TullyFisher1977} relation (see the left panel of Figure \ref{SMF}). We then use the best GPR fit to derive the stellar masses for each individual HIPASS galaxy from their \hi\ line widths.

The effective volume $V_{\rm eff}$ for each HIPASS galaxy is derived using a bicariate stepwise maximum likelihood technique \citep{Zwaan2004}. After binning the data, we sum the values of $\frac{1}{V_{\rm eff}}$ for galaxies within each bin following \citet{Zwaan2010}. This gives the stellar mass function. There are two sources for the uncertainties: one from the poisson distribution which is given by the square root of the summation of $V^{-2}_{\rm eff}$, and the other one from the scatter of the $W^i_{\rm P50}/2-m_\star$ relation. To account for the latter, we add Gaussian noise (the standard deviation of the Gaussian noise is given by the scatter of the $W^i_{\rm P50}/2-m_\star$ relation) to the estimated stellar mass for each HIPASS galaxy and measure a new SMF. After 10000 random iterations, we calculate the standard deviations of 10000 HMFs and add them to the poisson errors in quadrature.

The result is plotted in Figure \ref{SMF} together with the SMF measured by \citet{Moffett2016} from the Galaxy and Mass Assembly survey \citep{GAMA2015}. \citet{Moffett2016} measured the SMFs for different morphologies. Disk dominated galaxies contain most of the cold gas in galaxies, so make the most direct comparison to \hi-selected HIPASS galaxies. Figure \ref{SMF} shows a satisfactory agreement between these two measurements covering the available mass range. This confirms that our method can measure a mass function, and match one that is independently measured by a completely different type of survey.

\subsection{Halo Mass Function}
\label{sec:HMF}

Using the best GPR fits shown in Figure \ref{WP50M200}, we derived the halo masses of the HIPASS galaxies for the three profiles. Summing the values of $V^{-1}_{\rm eff}$ within each halo mass bin, we obtain the halo mass functions. We estimate the uncertainties using the same method as for the stellar mass function. 

\begin{figure*}[t]
\epsscale{1.17}
\includegraphics[scale=0.3]{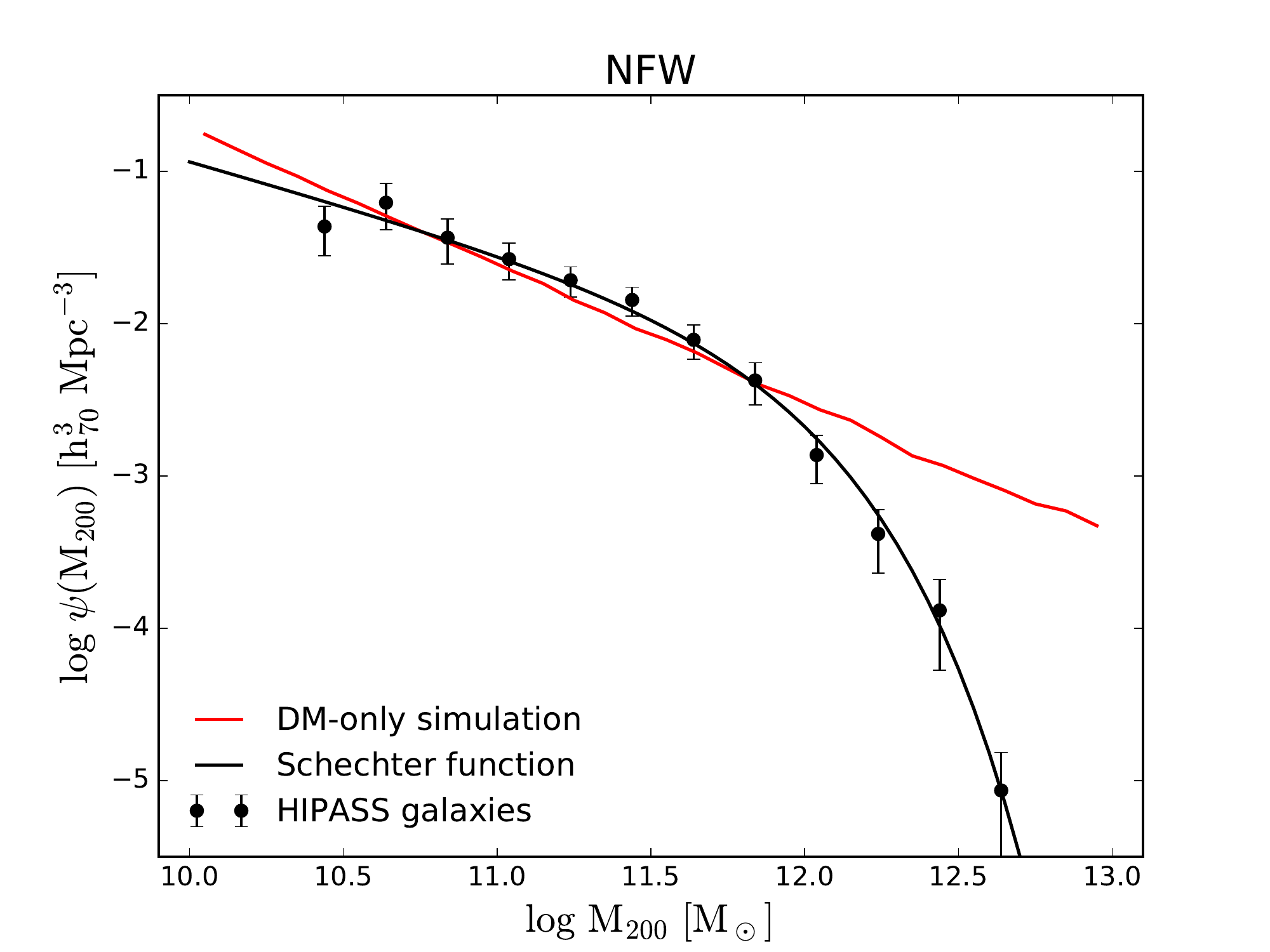}\includegraphics[scale=0.3]{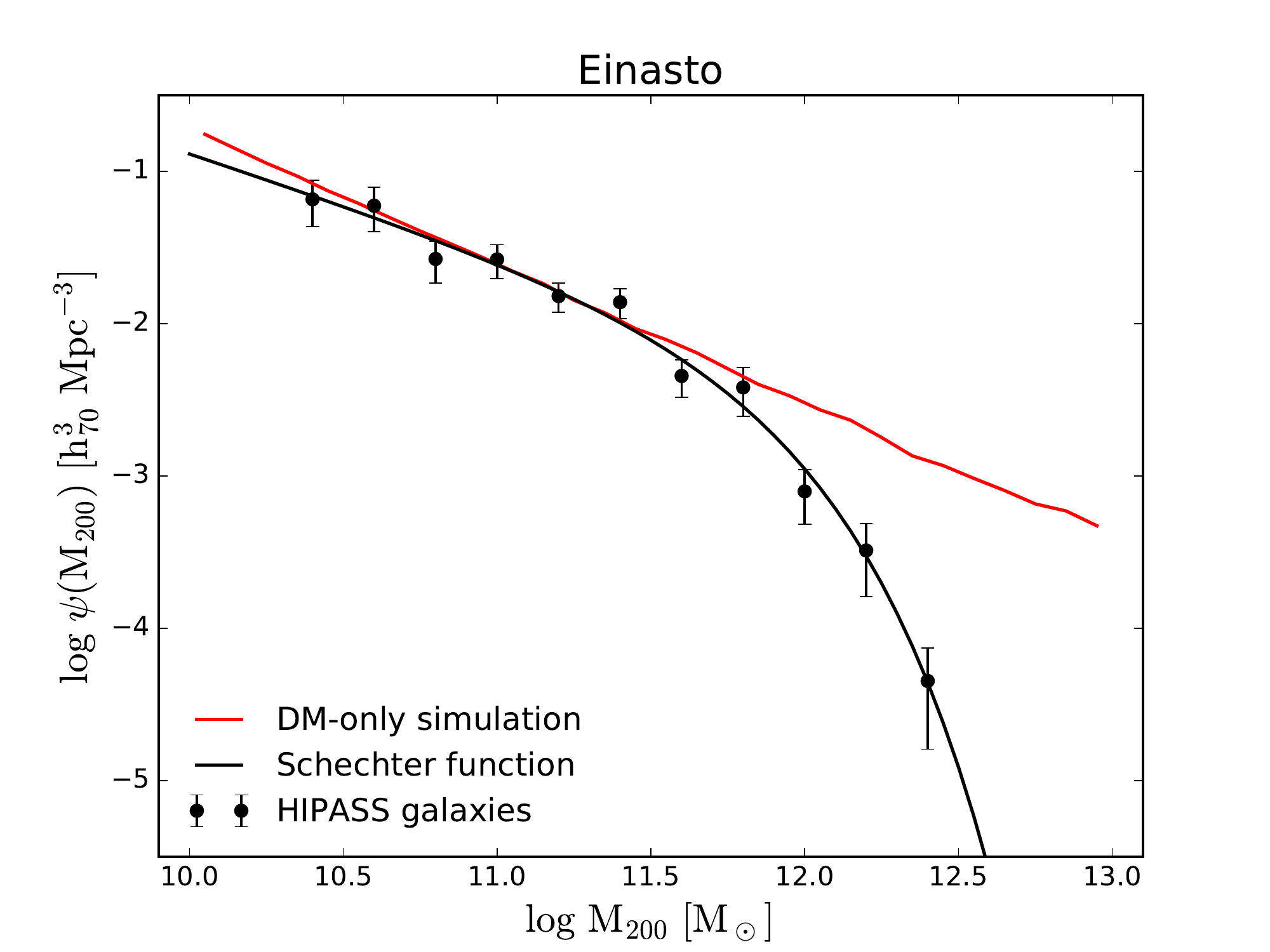}\includegraphics[scale=0.3]{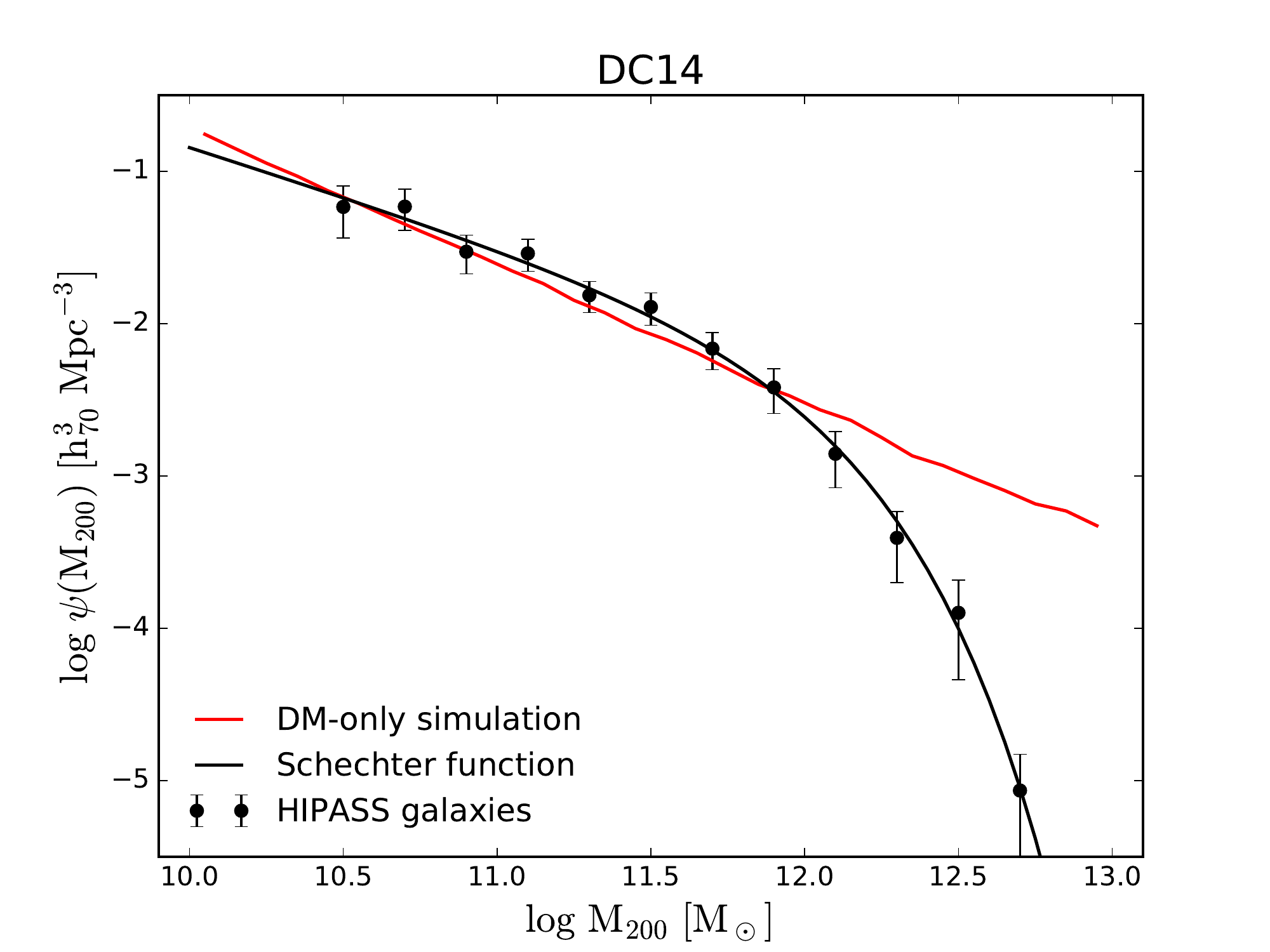}
\caption{The halo mass functions measured using the HIPASS galaxies \citep{Meyer2004, Zwaan2010} for the NFW (left), Einasto (middle) and DC14 (right) profiles. Solid black lines are the best-fit modified Schechter functions. Red lines represent the prediction of DM-only simulations \citep{Sprengel2018, Nelson2019}.
}
\label{HMF}
\end{figure*}

The HMFs for the NFW, Einasto, and DC14 profiles are shown in Figure \ref{HMF}. The bins are set to avoid being only partially covered by the data. They are similar in shape, given the similar $W^i_{\rm P50}/2-M_{200}$ correlations for the three halo profiles. The HMFs are well fit by the modified Schechter function,
\begin{equation}
\psi(M_{200}) = \psi_\star\Big(\frac{M_{200}}{M_\star}\Big)^{\alpha+1}\exp\big(-\frac{M_{200}}{M_\star}\big)\ln10.
\end{equation}
The corresponding parameters are listed in Table \ref{Schechter}. 

The integral of the Schechter function gives the mass density of DM associated with galaxies detected in \hi:
\begin{equation}
\rho_{\rm DM} = \psi_\star M_\star[\Gamma(\alpha+2, \frac{M_{\rm up}}{M_\star}) - \Gamma(\alpha+2, \frac{M_{\rm low}}{M_\star})],
\end{equation}
where $\Gamma(\alpha+2, x) = \int^x_0x^{\alpha+1}e^{-x} {\rm d}x$ is the incomplete Gamma function, and $M_{\rm up}$ and $M_{\rm low}$ are the upper and lower limits of the integrating masses, respectively. We calculate $\Omega_{\rm m,gal}=\rho_{\rm DM}/\rho_{\rm crit}$ in the mass range between 10$^{10.5}$ and 10$^{12.5}$ M$_\odot$. We find that the DM mass density in \hi-detected galaxies is only about a tenth of the cosmic DM density in the probed mass range, as shown in Table \ref{Schechter}. Even if we integrate the best-fit Schechter function from zero to infinity, the DM mass density is still smaller than 0.04. This suggests that most DM in the universe is not bound to \hi-rich galaxies.

\begin{deluxetable}{lcccc}
\tablenum{1}
\tablecaption{The best-fit parameters of the modified Schechter function for the NFW, Einasto and DC14 profiles. $\Omega_{\rm m, gal}$ is the integrated dark-matter mass density. \label{Schechter}}
\tablewidth{0pt}
\tablehead{
\colhead{Model} & \colhead{$\psi_\star\times10^3$}  & \colhead{$\log M_\star/M_\odot$} & \colhead{$\alpha$} & \colhead{$\Omega_{\rm m,gal}$}
}
\startdata
NFW & 4.44 $\pm$  0.84 & 11.86 $\pm$  0.03 & -1.57 $\pm$ 0.08 & 0.031 \\
Einasto & 3.93 $\pm$ 1.09 & 11.76 $\pm$ 0.05 & -1.66 $\pm$ 0.10 & 0.023\\
DC14 & 3.60 $\pm$ 0.57 & 11.94 $\pm$ 0.02 & -1.64 $\pm$ 0.06 & 0.034\\
\enddata
\end{deluxetable}

The empirical HMF that we derive is in reasonable agreement with theoretical expectations from $\Lambda$CDM for all halo types (Fig.\ \ref{HMF}). This holds at intermediate and low halo masses down to $\sim 10^{10.5}$ M$_\odot$. Galaxies with lower masses are generally not observed in current, single-dish surveys \citep{Papastergis2013, Guo2017} and hence are missing in the HIPASS sample.

A halo mass of $10^{10.5}$ M$_\odot$ corresponds to a stellar mass of $\sim 10^{8}$ M$_\odot$. This is typical of low-mass dwarf Irregulars in the field, which are usually gas rich, often having more gas than stars \citep{LSBSFMS}. Consequently, this stellar mass may correspond to a wide range of baryonic masses (the sum of stars and gas). Though low mass, these galaxies are more massive than the satellite galaxies of the Local Group. Consequently, we may not have reached the regime where the missing satellite problem takes hold \citep{Tikhonov2009, Bullock2017}.

At high masses, the VF of HIPASS galaxies truncates sharply above $W_{\rm P50}>200$ km/s \citep[Figure 1 of][]{Zwaan2010}. Consequently, our empirical HMF shows a corresponding cut-off above $M_{200} = 10^{12}$ $M_{\odot}$, comparable to the mass of the Milky Way. Intriguingly, the ALFALFA survey finds more high-widths galaxies than HIPASS and its VF truncates at slightly larger values of $W_{\rm P50}>300$ km/s \citep{Papastergis2011}. Thus, the ALFALFA data must still imply a cut-off in the empirical HMF, albeit at slightly larger halo masses. This may seem problematic compared to the predicted halo mass function, which continues as a power law. However, the sharp cut-off in the observed HMF does not preclude the existence of more massive halos, provided that they are \hi\ poor. Early-type galaxies fit this description, and fill out the top end of the stellar mass function in Fig.\ \ref{SMF}. Further tests will require careful interrogation of hydrodynamical simulations that select mock galaxies in a way that matches the HIPASS survey. This is beyond the scope of the present work, so it remains an open question whether the current generation of simulations is consistent with these observations.

\section{Discussion and conclusion}
\label{sec:disc}

In this paper, we present an empirical method to derive the halo mass function of galaxies. We first determine the correlation between \hi\ line width and DM halo mass as determined from rotation curve fits utilizing the NFW, Einasto, and DC14 halo models. We use this correlation to assign halo masses to galaxies detected in the HIPASS \hi\ survey. It is then possible to map the observed velocity function to the actual halo mass function.

We detect no analog to the missing satellite problem down to a halo mass of $10^{10.5}$ M$_\odot$. However, our halo mass function only spans two dex in halo mass compared with the much larger range in the stellar mass function. This is due to the nonlinear stellar mass--halo mass relation \citep[see][]{Moster2013}. It suggests that
\begin{equation}
\log M_\star \propto (\beta+1)\log M_{\rm 200},
\end{equation}
at $M_{200}<M_1 = 10^{11.59}$ M$_\odot$, where $\beta=1.376$. As such, if the HIPASS galaxies span 4 dex in stellar mass, their halo masses span only $4/(\beta+1) = 1.7$ dex. This nonlinearity compresses an approximately flat observed VF \citep{Zwaan2010} into a less extended, more steeply rising HMF.

The stellar mass--halo mass relation of abundance matching was imposed as a prior in fitting the SPARC rotation curves. On the one hand, this is appropriate to the extent that abundance matching has become an essential aspect of the $\Lambda$CDM paradigm. On the other hand, the correlation between halo mass and \hi\ line width is less clear if we do not impose the stellar mass--halo mass relation as a prior. If instead we were to make the natural assumption that $M_{200} \sim W_{50}^3$ \citep{Post2019}, the low-mass end of the HMF would be shallower than predicted. Abundance matching thus plays a key role in reproducing the predicted halo abundance at intermediate and low halo mass.

Accepting the abundance-matching prior on halo masses obtained from rotation curve fits, we find good agreement between the predicted and measured halo mass functions at intermediate and low halo masses down to $10^{10.5}$ M$_\odot$. 
Below this mass limit, there is a hint of a discrepancy in the field analogous to the missing satellite problem. 
To explore if this is a genuine problem requires pressing the mass limit of blind \hi\ surveys to lower masses. This will be possible with large interferometric \hi\ surveys with the SKA and its pathfinders. 

\acknowledgments

We thank Yong Tian, Harry Desmond, and Harley Katz for useful discussions. This work was supported in part by NASA ADAP grant 80NSSC19k0570.

\bibliographystyle{aasjournal}
\bibliography{PLi}

\end{document}